\documentclass[conference]{IEEEtran}

\usepackage{cite}
\usepackage{amsmath,amssymb,amsfonts}
\usepackage{algorithmic}
\usepackage{graphicx}
\usepackage{textcomp}
\usepackage{xcolor}
\usepackage[nolist]{acronym}

\usepackage{hyperref}

\hypersetup{
	hidelinks,
	bookmarks=false,
	pdftitle={DQN-AF: Deep Q-Network based Adaptive Forwarding Strategy for Named Data Networking},
	pdfauthor={Ygor Amaral B. L. de Sena and Kelvin Lopes Dias and Cleber Zanchettin},
	pdfsubject={Accepted to be published in: 2020 IEEE LATINCOM},
	pdfcreator={TeX},
	pdfkeywords={Fowarding Strategy, Deep Reinforcement Learning, DQN, Named Data Networking.}
}

\def\BibTeX{{\rm B\kern-.05em{\sc i\kern-.025em b}\kern-.08em
    T\kern-.1667em\lower.7ex\hbox{E}\kern-.125emX}}

\begin{document}

\newcommand\copyrighttext{ 
	\Huge {IEEE Copyright Notice} \\ \\
	\large {\textcopyright~2020 IEEE. Personal use of this material is permitted. Permission from IEEE must be obtained for all other uses, in any current or future media, including reprinting/republishing this material for advertising or promotional purposes, creating new collective works, for resale or redistribution to servers or lists, or reuse of any copyrighted component of this work in other works.} \\ \\

	Accepted to be published in:
	\begin{itemize}
		\item Proceedings of the 2020 IEEE Latin-American Conference on Communications (IEEE LATINCOM 2020), Nov 18-20, 2020.
	\end{itemize}
}

\twocolumn[
\begin{@twocolumnfalse}
	\copyrighttext
\end{@twocolumnfalse}
]

\title{DQN-AF: Deep Q-Network based Adaptive Forwarding Strategy for Named Data Networking}

\author{Ygor Amaral B. L. de Sena\textsuperscript{$\mp \star$}, Kelvin Lopes Dias\textsuperscript{$\mp$}, Cleber Zanchettin\textsuperscript{$\mp$}\\
	\textsuperscript{$\mp$}Centro de Informática, Universidade Federal de Pernambuco, (CIn/UFPE), Brazil\\
	\textsuperscript{$\star$}Unidade Acad\^emica de Serra Talhada, Universidade Federal Rural de Pernambuco, (UAST/UFRPE), Brazil\\
	ygor.amaral@ufrpe.br, \{kld, cz\}@cin.ufpe.br}

\begin{acronym}[ACRONYM] 
	\acro{ASF}[ASF]{\emph{Adaptive Smoothed RTT-based Forwarding}}
	\acro{CS}[CS]{\emph{Content Store}}
	\acro{DQ-Learning}[DQ-Learning]{\emph{Data-based Q-Learning}}
	\acro{DQN}[DQN]{\emph{Deep Q-Network}}
	\acro{DQN-AF}[DQN-AF]{\emph{DQN-Adaptive Forwarding}}
	\acro{FIB}[FIB]{\emph{Forwarding Information Base}}
	\acro{IQ-Learning}[IQ-Learning]{\emph{Interest-based Q-Learning}}
	\acro{NACK}[NACK]{\emph{Negative-Acknowledgement}}
	\acro{NDN}[NDN]{\emph{Named Data Networking}}
	\acro{NS}[NS]{\emph{Network Simulator}}
	\acro{NFD}[NFD]{\emph{NDN Forwarding Daemon}}
	\acro{PIT}[PIT]{\emph{Pending Interest Table}}
	\acro{ReLU}[ReLU]{\emph{Rectified Linear Unit}}
	\acro{RTT}[RTT]{\emph{Round Trip Time}}
	\acro{RTO}[RTO]{\emph{Retransmission Timeout}}
	\acro{SRTT}[SRTT]{\emph{Smoothed RTT}}

\end{acronym}

\maketitle

\begin{abstract}
\ac{NDN} has gained significant attention due to the appearance of several unforeseen design flaws that became evident with new communication scenarios. Among its many features, the two standard NDN forwarding strategies are not adaptive, causing performance degradation in several scenarios. This paper proposes an adaptive forwarding strategy based on deep reinforcement learning with Deep Q-Network, which analyzes the NDN router interface metrics without creating signaling overhead or harming the design principles from the NDN architecture, besides showing significant performance gains compared to the standard strategies.
\end{abstract}


\section{Introduction}\label{sec:introduction}
With massively connected devices and the popularization of new applications (e.g., high-quality video streaming), the future Internet design will have new requirements, such as enhanced device mobility. The \ac{NDN}~\cite{Zhang2014} is a promising internet architecture paradigm that has been proposed to mitigate the disadvantages of the original Internet architecture. Unlike traditional IP networks, \ac{NDN}s removes the need for end-to-end communication, with a new approach to distributing and retrieving content on the Internet.

As \ac{NDN} focuses on content-driven communication and content identification, its operation mode has become quite similar to the publish-subscribe communication pattern, where consumers request content through interest packets and producers respond with data packets. In this context, three data structures are required: \ac{CS}, \ac{PIT}, and \ac{FIB}, in addition to the forwarding strategy module~\cite{Zhang2014}.

An important feature of the \ac{NDN} architecture is that due to \ac{PIT}, which awaits the arrival of data packets, \ac{NDN} routers maintain the state of pending interests. Each interest has a lifetime and its respective registration to the \ac{PIT} is removed due to a timeout event, so that, among other metrics, routers can measure the packet delivery performance~\cite{Yi2012}, and how to calculate the \ac{RTT}. Moreover, the \ac{FIB} information is inserted by a routing algorithm, such as NLSR~\cite{Hoque2013} or manually by the network administrator, however, it is the forwarding strategy module, which decides whether, how and to which next hop an interest packet must be sent.

In the standard \ac{NDN} proposal, the two main strategies are currently based on route (best-route)~\cite{Yi2012} and flooding (multicast), with the first being strongly dependent on routing policies and prone to inefficiency when there are multiple available next-hops. The latter is considered costly, as it sends interest packets to all the next available hops, causing great network overload. For this reason, other adaptive forwarding strategies have been proposed, which better observe the network status to decide upon which next-hops the interest packet should be sent to. The most promising approaches are the \ac{ASF}~\cite{Lehman2016a}, which chooses the best next-hop based on \ac{SRTT} measurements and \ac{DQ-Learning}~\cite{Fu2017} that exchanges information between \ac{NDN} nodes using reinforcement learning algorithm to modify data packets to choose the best next hop.

In general, the interest forwarding strategies are still an open issue due to the remarkable drawbacks on the previously developed approaches. This paper proposes \ac{DQN-AF}, an adaptive forwarding strategy based on deep reinforcement learning with \ac{DQN}~\cite{Mnih2013} for \ac{NDN} networks. A prototype for the \ac{NFD}~\cite{Afanasyev2014} was developed using TensorFlow~\cite{Abadi2016} and simulated in {{ndnSIM}}~\cite{Mastorakis2017}. The results obtained in the simulations show the effectiveness of the proposal, which presents a superior performance compared to Best-Route and \ac{ASF}, in addition to achieving similar indexes compared to \ac{DQ-Learning}, with the advantage of not modifying data packets.

The paper is organized as follows: Section~\ref{sec:related_works} presents the related work. Section~\ref{sec:proposal} describes the proposed \ac{NDN} adaptive forwarding strategy. Section~\ref{sec:experiments} details the experimental setup. In Section~\ref{sec:results} numerical results are presented. Section~\ref{sec:conclusions} concludes the work.

\section{Related Work}\label{sec:related_works}
The works~\cite{Yi2012} and~\cite{Yi2013} describe the main characteristics of the \ac{NDN} architecture, which allow routers to evaluate the performance of the interfaces and thus quickly detect communication failures, helping to discover the most appropriate instant to try alternative paths in order to make adaptive interest packet forwarding. Even so, the main strategy of the standard \ac{NDN} architecture named Best-Route still lacks adaptive characteristics, which allows underutilizing \ac{NDN} router potential.

When interest packets arrive, the Best-Route strategy will always check which next hop is best classified in the \ac{FIB}, even if the link for this hop is experiencing communication problems, and the routing algorithm still has not updated the \ac{FIB}. This strategy will only try an alternative path when the \ac{FIB} is updated, either by disabling that hop's entry or by worsening its position in the classification. The problem is that routing algorithms do not usually perform updates as often or at an optimal time.

The \ac{ASF}~\cite{Lehman2016a} is an adaptive forwarding strategy that periodically performs probes all interfaces to measure the respective \ac{SRTT}. From this metric, \ac{ASF} sorts the interfaces to choose the best next-hop and whenever one interface suffers a timeout, it is penalized and falls to the end of the classification. If two interfaces have the same \ac{SRTT} value, \ac{ASF} will behave similarly to Best-Route, choosing the interface with the best position in the \ac{FIB} classification.

\ac{DQ-Learning}~\cite{Fu2017} is another adaptive strategy. It is based on the Q-Learning reinforcement learning algorithm, which will update the Q value through positive and negative rewards upon receiving data packets. The Q value represents the delay for the next hop, thus lower Q values reflect better interfaces. In addition, when a loss is detected, Q will increase in value. The problem regarding this approach relies on the principle of data immutability that is not respected since \ac{NDN} nodes will modify each data packet before sending, to add the following information: the best Q value (zero if producer node) and the sending time, which is information used to update the Q value on the node that will receive the data packet. In this way, the \ac{NDN} nodes propagate the network status so that each router can decide its next hops. If the same interface has successive losses, DQ-Learning avoids penalizing its respective Q value unnecessarily by temporarily interrupting the interface choice by Q-Learning and using probability theory. The authors in~\cite{Fu2017} also proposed the \ac{IQ-Learning} approach, which works similarly to \ac{DQ-Learning}, but instead of modifying the data packet to provide the algorithm information, this approach creates a new packet (non-standard) for each feedback, causing extra signaling overhead. The authors themselves claim that the performance of \ac{DQ-Learning} is superior to \ac{IQ-Learning}.

Finally, regular Multicast is another strategy that belongs to the \ac{NDN} standard architecture. In this approach, upon receiving an interest packet to be sent, it will greedily replicate all the next hops interest. An advantage is that it increases the chances of receiving the respective data packet; however, at a very high cost for the network, as can be seen in the results presented in Section~\ref{sec:results}. 

\section{Adaptive Forwarding Strategy}\label{sec:proposal}
\begin{figure}[htbp]
	\centerline{\includegraphics[width=.4\textwidth]{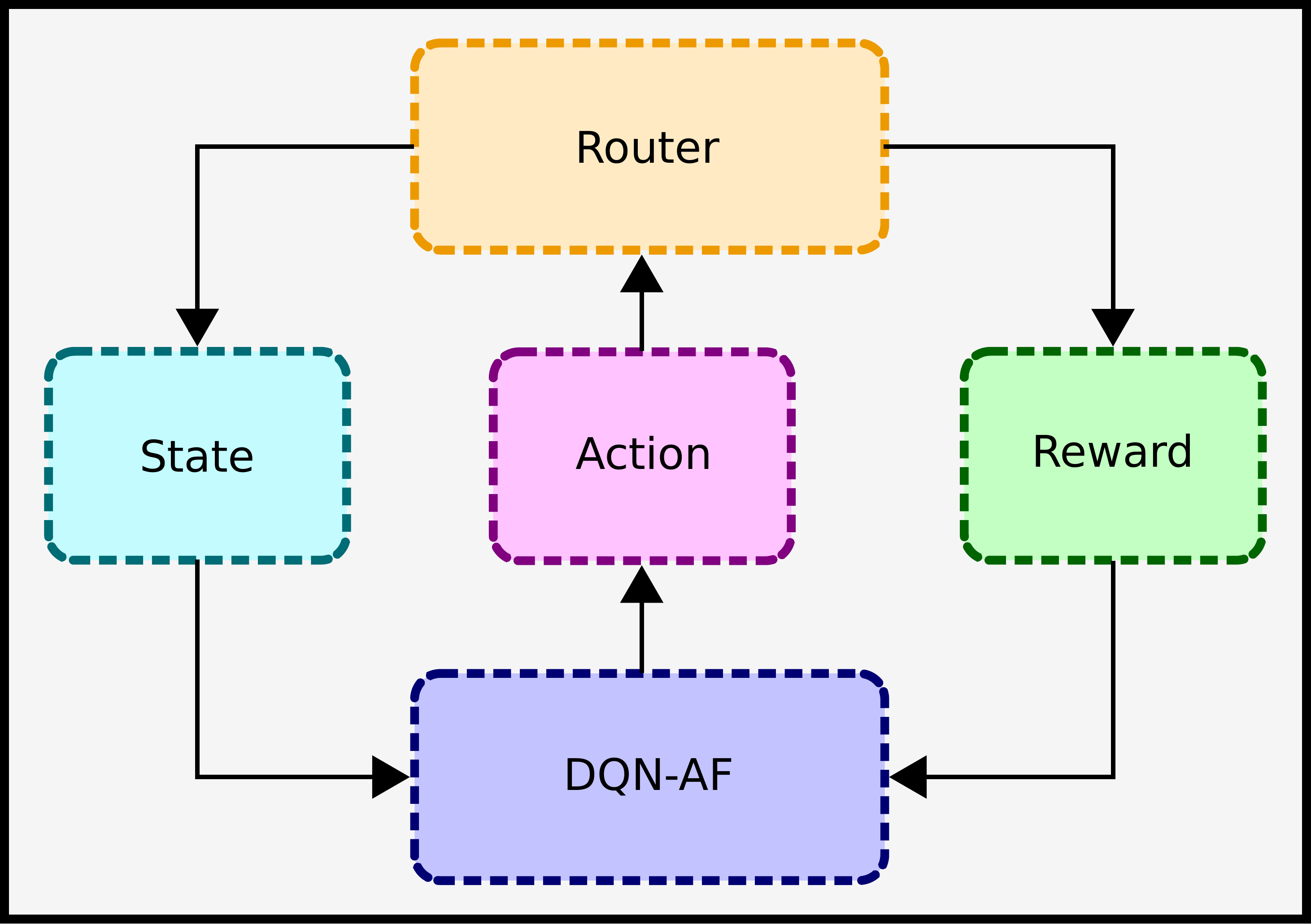}}
	\caption{Adaptive forwarding operation overview with \ac{DQN-AF}.}
	\label{fig:overview_dqn_af}
\end{figure}

This section describes the adaptive forwarding operation depicted in Fig.~\ref{fig:overview_dqn_af}. The workflow starts with the router collecting data from available links (known from the \ac{FIB}) such as numbers of packets delivered and lost, \ac{RTT}, \ac{SRTT}, \ac{RTO}. The collected data is used to identify the environment states and be processed by the \ac{DQN-AF} agent. Whenever an interest packet arrives at the \ac{PIT}, the \ac{NDN} router informs the states that the \ac{DQN-AF} agent can predict which interface should be the next hop. The agent then shares the router's decision, which sends the interest to the selected interface and creates a record at the \ac{PIT}. The \ac{NDN} router will only inform the reward in two possible situations: (a) upon respective data packet arrival or (b) loss detection, by \ac{NACK} or timeout. If the data packet successfully arrives, a positive reward will be reported to the \ac{DQN-AF}, whereas if a loss is detected, a negative reward is generated instead. In this way, the proposed forwarding strategy can adapt to different network conditions.

\subsection{DQN-Adaptive Forwarding (DQN-AF)} \label{sec:details_dqn_af}

In this paper, we take advantage of the fact that \ac{NDN} routers maintain state information of sent requests to collect performance metrics from the router’s interfaces, allowing the \ac{DQN-AF} module to generate state data sent to the deep reinforcement learning agent. In the current version, \ac{DQN-AF} measures the following metrics on each interface:

\begin{itemize}
	\item $\eta \rightarrow$ number of packets sent over a time interval $t$;
	\item $\omega \rightarrow$ rate of successful deliveries in the last $\eta$ interest packets sent with defined status (delivered or lost);
	\item $\delta \rightarrow$ stores if the interface is available and informs that a timeout occurred on the last sent interest.
\end{itemize}

In \ac{DQN-AF}, a state ($s$) is composed by $\omega$ and $\delta$ for each interface, so the size of the state is dependent on the number of interfaces. With these state values, the deep reinforcement learning algorithm can predict by observing data regarding the previous experiences (success/failures). This was defined because an interface with low success rate over time, but in the last verified interval delivered all packets, tends to be a better choice than an interface with high success rate, but that in the last interval delivered a small fraction of the packets.

\begin{figure}[htbp]
	\centerline{\includegraphics[width=.45\textwidth]{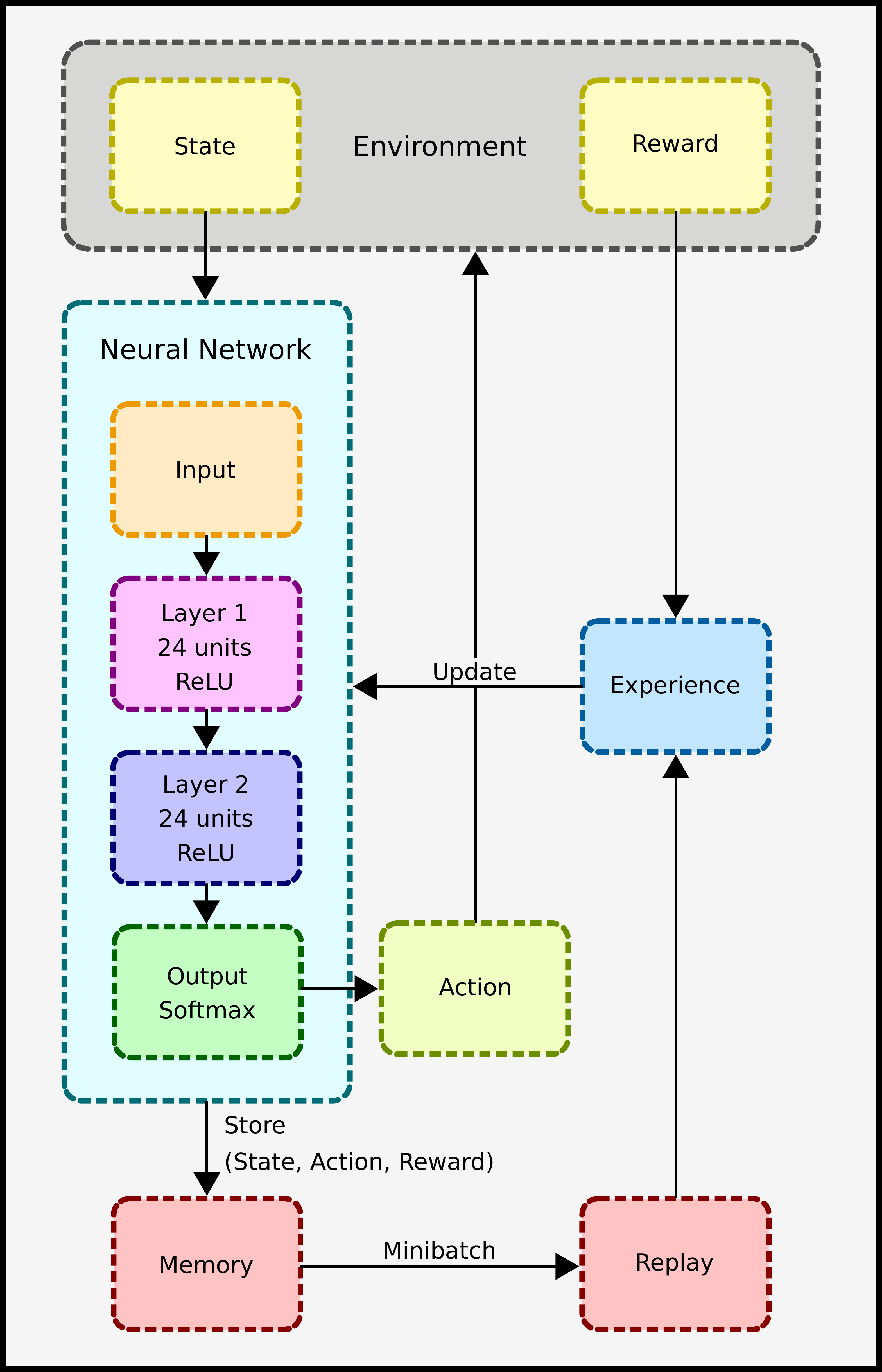}}
	\caption{Internal view of the \ac{DQN-AF}.}
	\label{fig:dqn_af_architecture}
\end{figure}

Fig.~\ref{fig:dqn_af_architecture} shows the details of the \ac{DQN-AF} strategy. When states are sent to the agent, data enters a fully connected multilayer perceptron neural network, with two hidden layers containing 24 units each and the \ac{ReLU} as the activation function. The number of available interfaces defines the neural network output size, and the unit values are generated by the Softmax activation function, which represents the Q values for each interface. The action ($a$) is formed from the best Q found in the output.

After the router acts as determined by the \ac{DQN-AF}, it waits until the status of the action is defined. When successful, a positive reward is generated with $SRTT / \omega$ from the interface and sent to the agent. In this way, this positive reward takes both the delay and the success rate; in this context, lower numerical values are obtained with higher rewards. When the action is unsuccessful, a negative reward is generated with an arbitrarily large value.

When the reward ($r$) arrives at the \ac{DQN-AF} agent, the neural network trains the weights with expression (\ref{eq:q_value}), to learn from the new experience (state, action, reward), where $\gamma$ is the discount factor, and $Q(s, a)$ is the prediction of a future reward for the action taken ($a$) from the state ($s$). 

\begin{equation}
	Experience = (1 - \gamma) * Q(s, a) + \gamma * r \label{eq:q_value}
\end{equation}

One of the main virtues of \ac{DQN} compared to Q-Learning is the ability to improve learning through repetition of past non-consecutive experiences, since learning only with consecutive samples is inefficient due to strong correlation between samples~\cite{Mnih2013}. Therefore, the \ac{DQN-AF} uses a memory of maximum size $\kappa$ to store the last analyzed experiences. For every $\rho$ analyzed experiments, the \ac{DQN-AF} will re-train experiences of a minibatch formed by randomly selected $\psi$ samples.

\section{Experiments} \label{sec:experiments}
The performance evaluation using the official tools of the \ac{NDN} project required developing a prototype for the forwarding strategy, which was composed of two main blocks: The first is a C++ module within the \ac{NFD} 0.6.5 (project router) that performs multiple tasks related to the adopted strategy such as link data collection, sending the interest to the next hop, and reward calculation. The second part implements \ac{DQN} and was developed in Python using TensorFlow 2.1.0 to use hardware acceleration through parallel processing boards. 

The following lines describe the \ac{DQN-AF} parameters. First, a random variable (between 0.0 and 1.0) is assigned to the possible actions. If the generated number is smaller than or equal to the $\epsilon$ (exploration rate), the action will be randomly defined. For each received reward, the $\epsilon$ decay by 0.005, until the minimum value 0.01 is reached. The $\gamma$ (discount factor) controls the weight of the future reward compared to the current reward. The value is set statically to 0.95. \ac{ReLU} was used in the hidden layers whereas the output layer uses Softmax as activation functions. The loss function was defined as the mean squared error. Multiple \ac{DQN-AF} variants have been tested with RMSprop and Adam (see Table~\ref{tab:settings_variants_dqn_af}) but RMSprop performed better (Section ~\ref{sec:results}), being the recommended optimizer. Three learning rates were evaluated (see Table ~\ref{tab:settings_variants_dqn_af}) but the value of 0.005 is recommended. When an operating cycle status--action--reward is ended (see in Fig.~\ref{fig:overview_dqn_af}), this experience is saved in memory, i.e., a queue that stores the last 2000 experiences is created. This parameter is named memory size ($\kappa$). For every 100 experiences analyzed ($\rho$), a minibatch is formed with 32 random samples selected from memory. This parameter is known as the minibatch size ($\psi$). This process re-trains experiences with low correlation in \ac{DQN}, reducing the forgetting effect. Lastly, a time interval ($t$) of 100 ms is used to measure $\eta$.

\subsection{Simulation Details}\label{sec:simulation_details}
The simulations were performed using the {{ndnSIM}}-2.7, the official \ac{NDN} project simulator based on \ac{NS} 3.29. To evaluate the \ac{DQN-AF} with different optimizers and learning rates, six variants were created (see Table ~\ref{tab:settings_variants_dqn_af}). Besides, each variant has been evaluated in two operation modes: a) without saving the weights, that is, with each new simulation performed, the \ac{DQN} is started with new weights; b) saving the weights, at the end of each simulation, the weights were saved so that in the next simulation the \ac{DQN} could reuse the previous values. The results are analyzed in Section~\ref{sec:results}.

\begin{table}[htbp]
	\caption{DQN-AF variants settings}
	\begin{center}
		\begin{tabular}{|c|c|c|}
			\hline
			\textbf{Variants}& \textbf{Learning rate} & \textbf{Optimizer} \\
			\hline
			dqn-af-1& 0.001 & \\
			\cline{1-2}
			dqn-af-2& 0.005 & RMSprop\\
			\cline{1-2}
			dqn-af-3& 0.01 & \\
			\hline
			dqn-af-4& 0.001 &  \\
			\cline{1-2}
			dqn-af-5& 0.005 & Adam\\
			\cline{1-2}
			dqn-af-6& 0.01 & \\
			\hline
		\end{tabular}
		\label{tab:settings_variants_dqn_af}
	\end{center}
\end{table}

In addition to \ac{DQN-AF}, the following strategies were also simulated: Best-Route, Multicast, \ac{ASF}, and \ac{DQ-Learning}. Except for the last, all others have an implementation available in {{ndnSIM}}. Hence, \ac{DQ-Learning} was entirely developed based on the parameters disclosed~\cite{Fu2017}.

\begin{figure}[htbp]
	\centerline{\includegraphics[width=.45\textwidth]{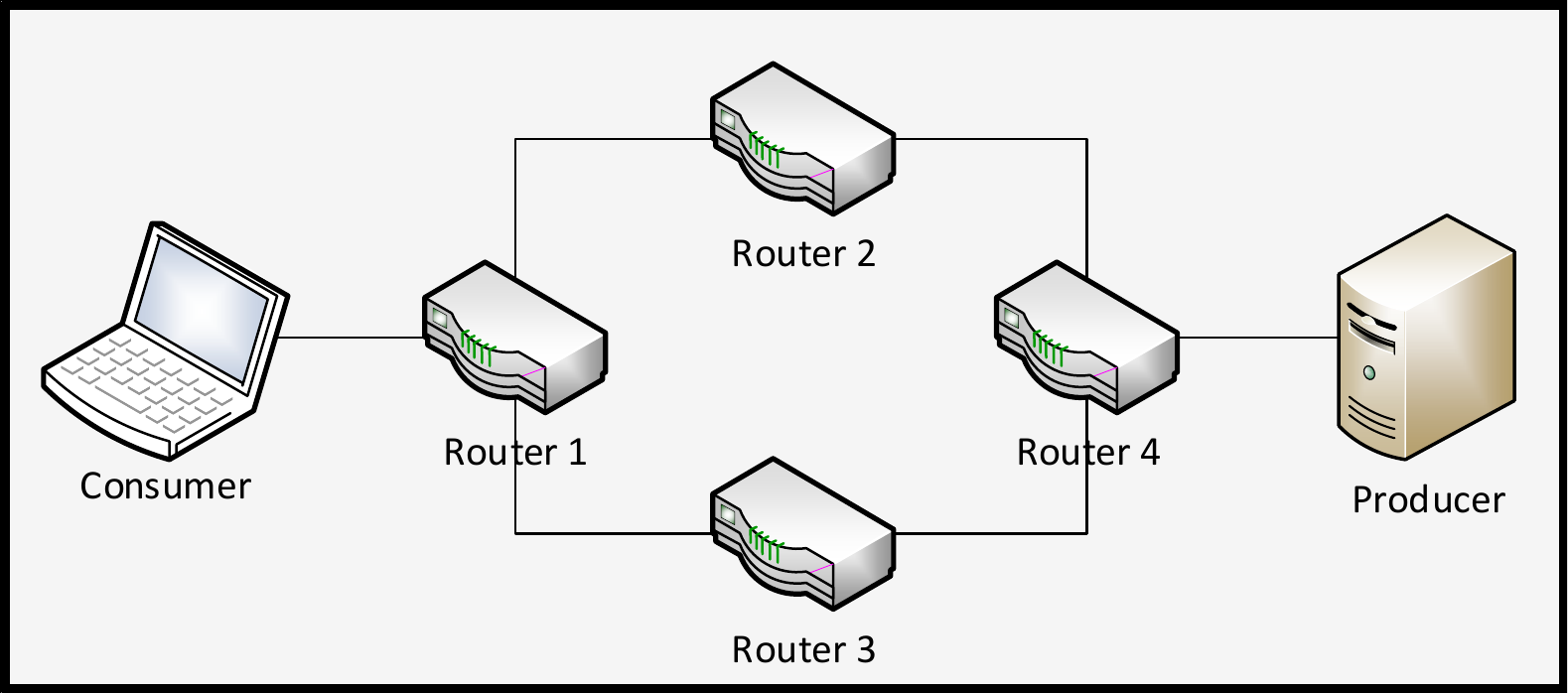}}
	\caption{Simulated NDN network topology.}
	\label{fig:topology}
\end{figure}

The adopted network topology (Fig.~\ref{fig:topology}) provides two paths between consumer and producer. The objective is to evaluate which forwarding strategy can deliver a greater amount of data packets. The evaluation focuses on router 1 since it can send interest packets for the two next possible hops. All links were configured with a 10 ms delay and a capacity of 1 Mbps. Each node in the network has a maximum queue of 10 packets. The consumer sends interests at a constant rate of 100 packets per second, and the producer responds to those interests by sending data packets with 1024 bytes of payload. The \ac{CS} cache was not used since the interests sent by the consumer are always for new data.

Each analyzed strategy and \ac{DQN-AF} variants were simulated 25 times reaching a total of 400 simulations with 30 seconds each (see Table~\ref{tab:settings_variants_dqn_af}). Furthermore, two performance metrics were collected: (1) Data received at the consumer -- Each data packet that arrives at the consumer is counted in order to check how far it was from the rate of 100 interest packets sent per second by the consumer. (2) Interests sent by router 1 -- This metric captures how many interests router 1 sent for its next hops. The aim is to compare the overload generated by the Multicast strategy with the other strategies.

Finally, to create an adverse situation, short-term instability events were generated in the links that connect router 1 to routers 2 and 3. The events are described below:

\begin{itemize}
	\item Event 1 (5.0s-9.0s) -- The link between router 1 and 3 loses all packets in that period.
	\item Event 2 (10.0s-14.0s) -- The link between router 1 and 2 loses all packets in that period.
	\item Event 3 (15.0s-19.0s) -- Again, the link between router 1 and 3 loses all packets in that period.
	\item Event 4 (20.0s-24.0s) -- Burst losses are generated in the links between router 1 and routers 2 and 3 with \textit{ns3::BurstErrorModel} configured at a rate of 0.02 and the average burst size in 10 packets following an exponential distribution.
	\item Event 5 (25.0s-29.0s) -- Bursts losses are repeated in the two links, only with a rate of 0.01.
\end{itemize}

\section{Results and Discussions}\label{sec:results}
\begin{figure}[htbp]
	\centerline{\includegraphics[width=.5\textwidth]{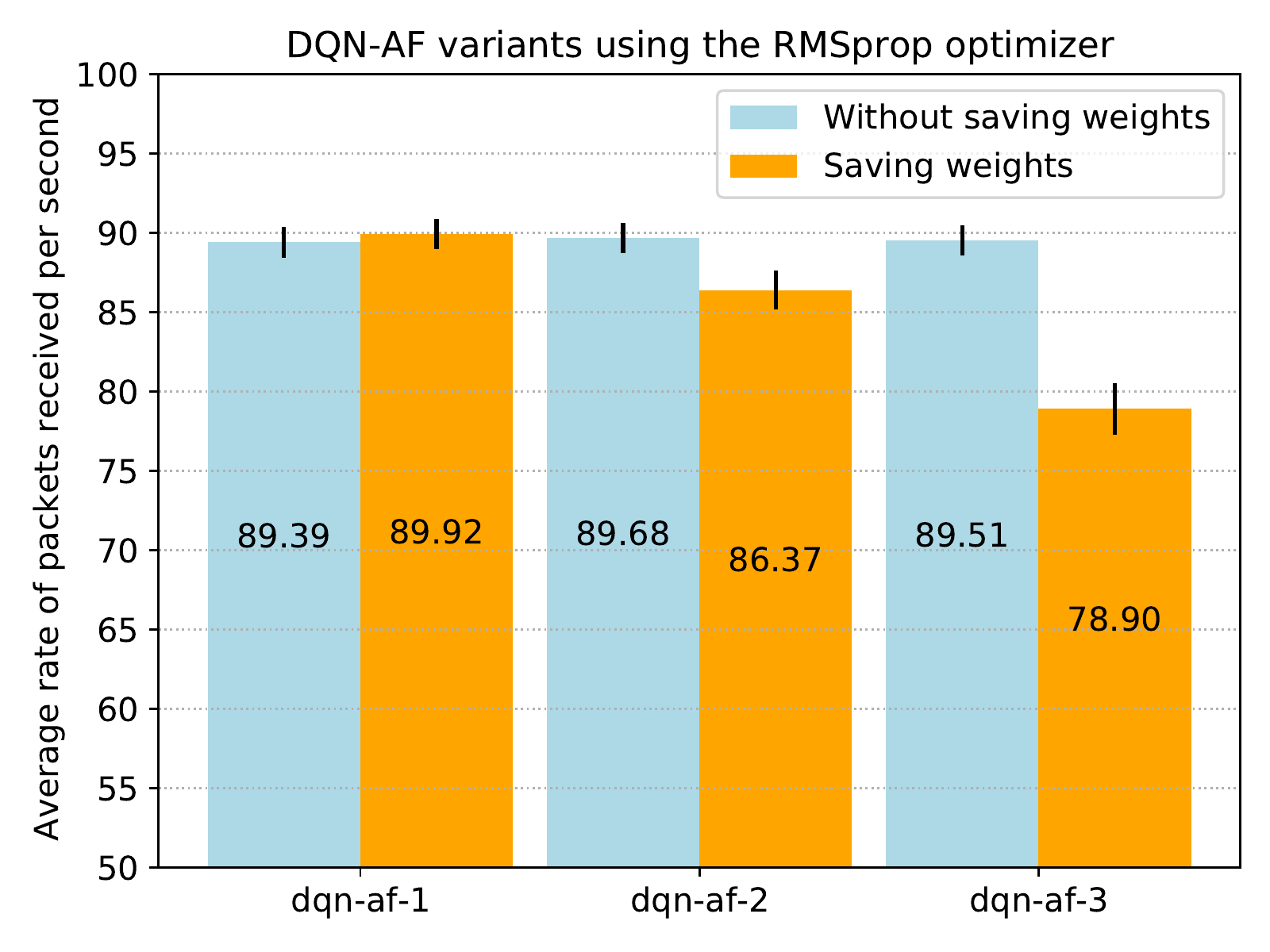}}
	\caption{Performance obtained by the DQN-AF variants configured with the RMSprop optimizer.}
	\label{fig:dqn-af-variants-rmsprop}
\end{figure}

\begin{figure}[htbp]
	\centerline{\includegraphics[width=.5\textwidth]{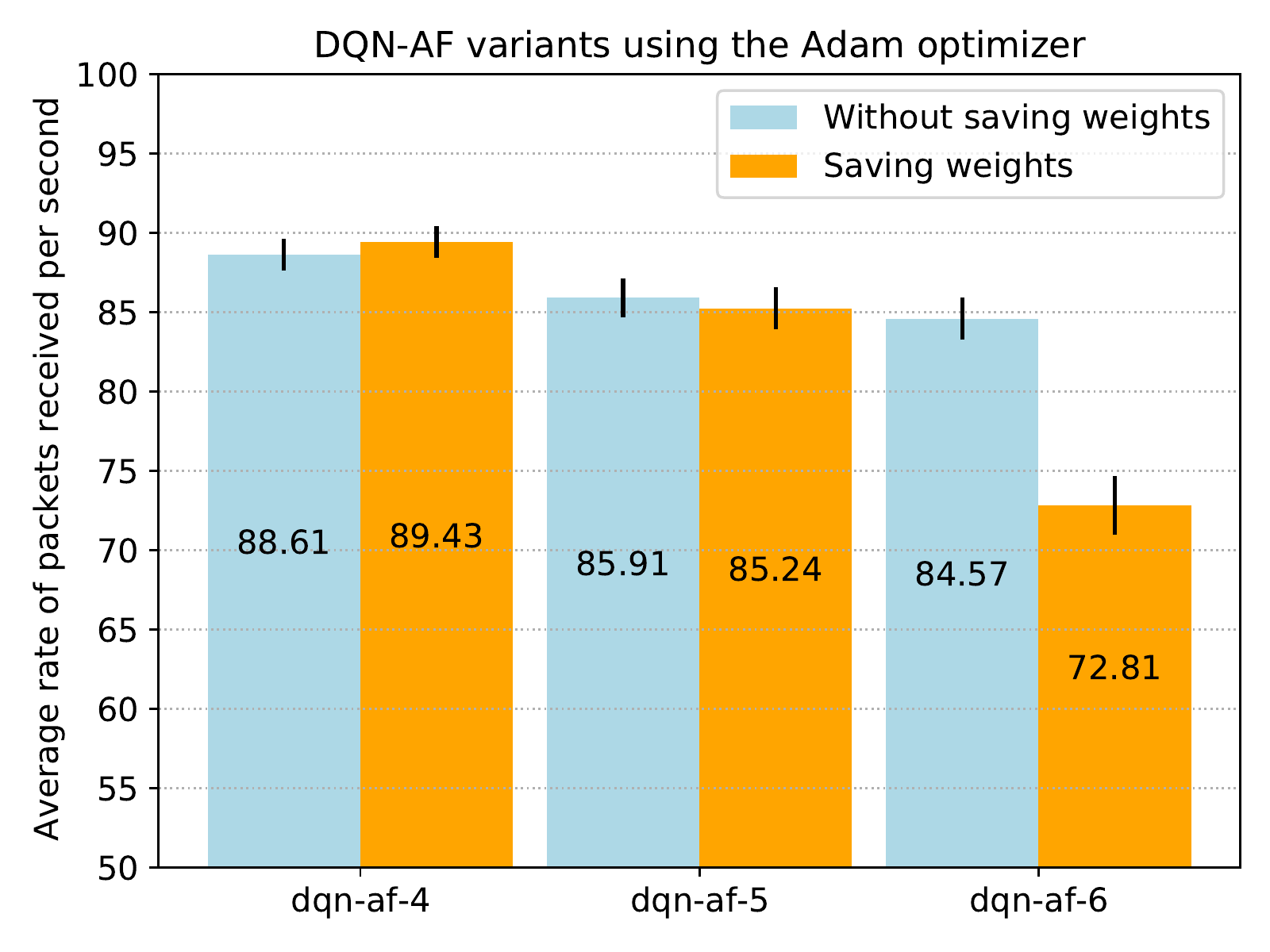}}
	\caption{Performance obtained by the DQN-AF variants configured with the Adam optimizer.}
	\label{fig:dqn-af-variants-adam}
\end{figure}

In this first evaluation, the six \ac{DQN-AF} variants (Table~\ref{tab:settings_variants_dqn_af}) were simulated. Each variant was tested in two operating modes: saving and without saving the neural network weights, as described in Section~\ref{sec:experiments}. The chart in Fig.~\ref{fig:dqn-af-variants-rmsprop} shows the average number of packets received by the consumer when testing different \ac{DQN-AF} variants that make use of RMSpro optimizer on router 1, while Fig.~\ref{fig:dqn-af-variants-adam} shows the variants with the Adam optimizer.

It is possible to observe that regardless of the adopted optimizer, keeping the weights between the simulations presented some advantage only when the learning rate was 0.001, in the other cases the performance worsened, being on average 11.76\% worse with \ac{DQN-AF}-6 when saving weights between the instances.

With regards to the optimizers, the RMSprop performed better in all cases, being around 0.49\% to 6.09\% higher than Adam. When analyzing different learning rates, the Adam optimizer is more sensitive towards this parameter than the RMSprop. The only satisfactory results using Adam were noted when the learning rate was 0.001.

The best results for the \ac{DQN-AF} were achieved with learning rates of 0.005 when the weights were not saved and 0.001 when saving them, always with the RMSprop optimizer. This may indicate that in scenarios where the \ac{DQN-AF} strategy will be used for a short period, the rate of 0.005 is indicated, otherwise, 0.001 should be adopted.

\begin{figure}[htbp]
	\centerline{\includegraphics[width=.5\textwidth]{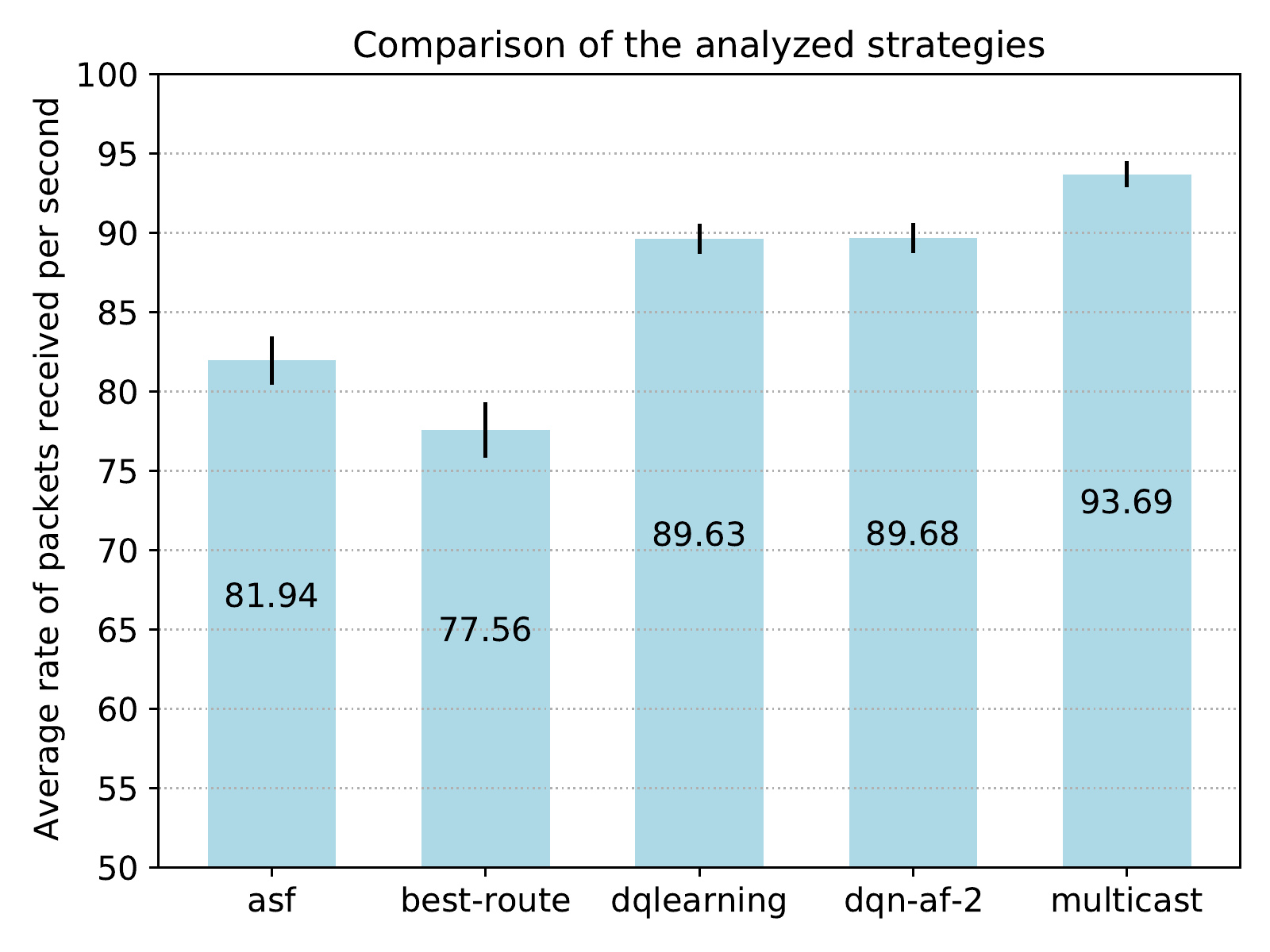}}
	\caption{Comparison between forwarding strategies with \ac{DQN-AF} (variant 2).}
	\label{fig:comparison-strategies}
\end{figure}

The second evaluation encompasses the comparison between Best-Route, Multicast, \ac{ASF} and \ac{DQ-Learning} with the \ac{DQN-AF}-02 variant (without saving weights). Fig.~\ref{fig:comparison-strategies} shows the average number of packets received by the consumer when testing the strategies analyzed in router 1. It is possible to observe that the strategy with the worst performance is Best-Route with an average of 77.56 received data packets per second.

Still, the Best-Route could have performed even worse, since the links between router 1 and routers 2 and 3 had the same characteristics, always presenting multiple ``best routes''. Thus, only in this case, the link between router 1 and router 2 was selected by order, considering that the routing information was saved in the \ac{FIB} data structure. Since the link between routers 1 and 2 was completely down for just 4 seconds versus 8 seconds of the link between routers 1 and 3, which was never used by the Best-Route strategy, as in these circumstances, it cannot be adaptive solution.

In general, the \ac{ASF} strategy performed better compared to the Best-Route approach but was worse compared to all others, with the consumer receiving, on average, 81.94 data packets per second. Although adaptive, the \ac{ASF} strategy checks only the \ac{SRTT} of each link to make a decision, and at different times the next two hops have similar \ac{SRTT}s, however, with one link delivering more packets than the other at that moment. In these moments, the \ac{ASF} ends up taking a little longer to understand which next jump is the best option.

\begin{figure}[htbp]
	\centerline{\includegraphics[width=.5\textwidth]{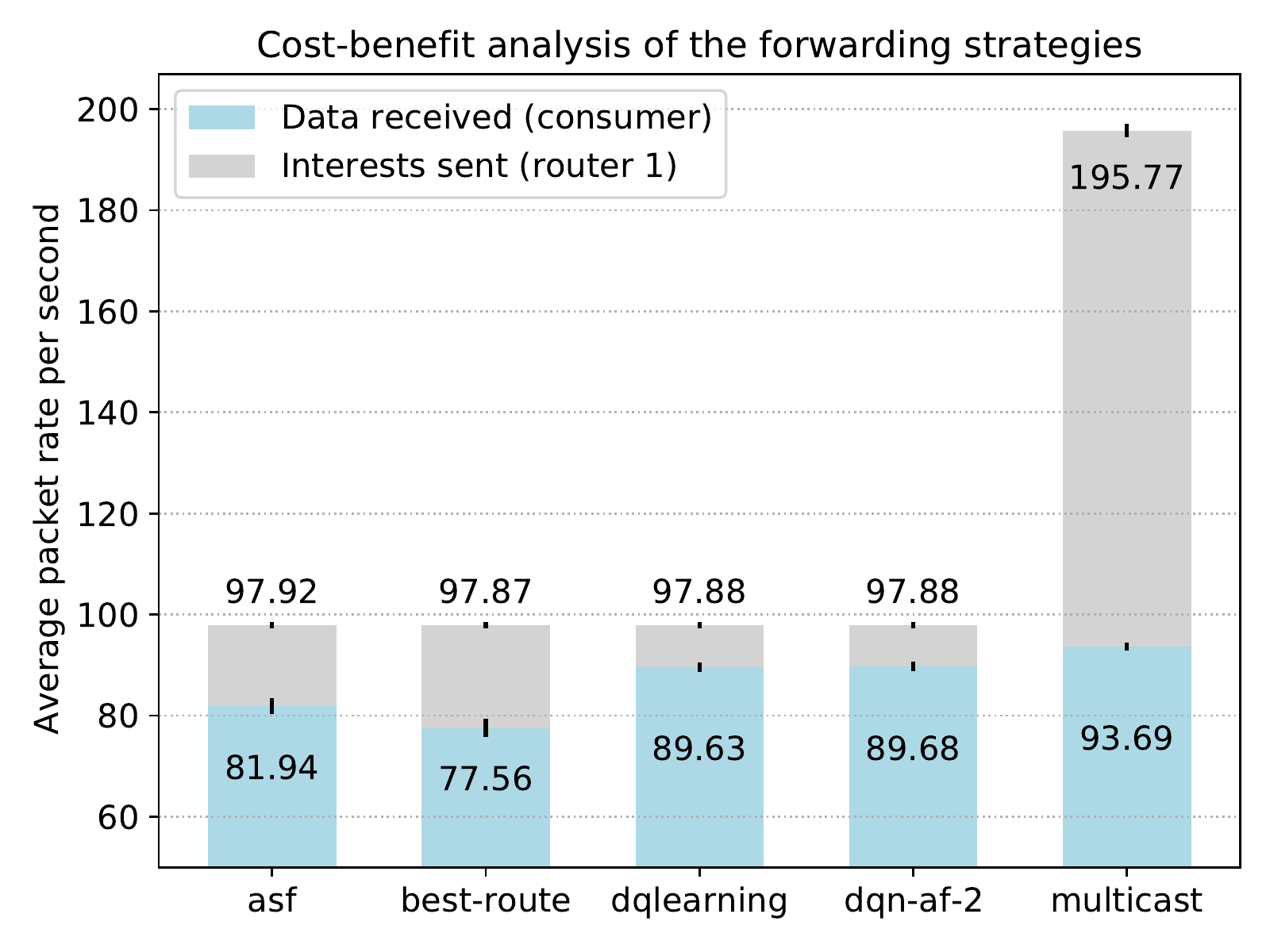}}
	\caption{Overload comparison with variant 2 of the \ac{DQN-AF}.}
	\label{fig:comparison-strategies-overload}
\end{figure}

The largest amount of data packets received at the consumer is on average 93.69 packets per second, which is achieved when the multicast strategy is used on router 1. However, this performance has a very high cost (see Fig.~\ref{fig:comparison-strategies-overload}), which presents the cost–benefit chart between the amount of interests sent by router 1 and the amount of data received at the consumer, with this, it is possible to evaluate the overload of each strategy. For achieving a rate of 93.69 packets, the Multicast strategy had to send on average 195.77 interests per second for the next two available hops, while all other strategies had a much lower overload, sending between 97.87 and 97.92 interests per second (on average).

Finally, the most cost--benefit strategies are \ac{DQ-Learning} and \ac{DQN-AF}-02 with respectively 89.63 and 89.69 data packets received per second (on average) and 97.88 sent interests per second for both. However, the \ac{DQ-Learning} operation mode requires \ac{NDN} nodes to modify data packet to add the time the packet was sent and the interface Q value that received the packet, so that routers have more information about the next hops. This characteristic of \ac{DQ-Learning} violates the design principle of data immutability, where data packets on \ac{NDN} networks must be immutable. On the other hand, our strategy presents a satisfactory performance with the advantage of not needing to modify packets, or even create unnecessary ones. \ac{NDN} nodes do not need to exchange information with each other, which does not generate any signaling overhead. The proposed \ac{DQN-AF} strategy only observes metrics of the interfaces of the router itself to predict which next hop is the most suitable to send the interest packets, without breaking any design principle of \ac{NDN} networks.

\section{Conclusions} \label{sec:conclusions}
In this paper we designed a comparative study considering four \ac{NDN} networks interest packets forwarding strategies and an adaptive forwarding alternative based on Deep Q-Network called \ac{DQN-AF} that selects the next hop of interest packets based on data from the links of the router itself, without needing any external information. Besides performing better than the standard NDN non-adaptive forwarding approaches, the proposed strategy also has the advantage of not breaking the principle of data immutability.

\bibliographystyle{bibliography/IEEEtran}
\bibliography{preprint-arXiv}

\begin{thebibliography}{10}
\providecommand{\url}[1]{#1}
\csname url@samestyle\endcsname
\providecommand{\newblock}{\relax}
\providecommand{\bibinfo}[2]{#2}
\providecommand{\BIBentrySTDinterwordspacing}{\spaceskip=0pt\relax}
\providecommand{\BIBentryALTinterwordstretchfactor}{4}
\providecommand{\BIBentryALTinterwordspacing}{\spaceskip=\fontdimen2\font plus
\BIBentryALTinterwordstretchfactor\fontdimen3\font minus
  \fontdimen4\font\relax}
\providecommand{\BIBforeignlanguage}[2]{{%
\expandafter\ifx\csname l@#1\endcsname\relax
\typeout{** WARNING: IEEEtran.bst: No hyphenation pattern has been}%
\typeout{** loaded for the language `#1'. Using the pattern for}%
\typeout{** the default language instead.}%
\else
\language=\csname l@#1\endcsname
\fi
#2}}
\providecommand{\BIBdecl}{\relax}
\BIBdecl

\bibitem{Zhang2014}
L.~Zhang, A.~Afanasyev, J.~Burke, V.~Jacobson, k.~claffy, P.~Crowley,
  C.~Papadopoulos, L.~Wang, and B.~Zhang, ``{Named Data Networking},''
  \emph{SIGCOMM Comput. Commun. Rev.}, vol.~44, no.~3, pp. 66--73, Jul. 2014.

\bibitem{Yi2012}
C.~Yi, A.~Afanasyev, L.~Wang, B.~Zhang, and L.~Zhang, ``{Adaptive Forwarding in
  Named Data Networking},'' \emph{SIGCOMM Comput. Commun. Rev.}, vol.~42,
  no.~3, pp. 62--67, Jun. 2012.

\bibitem{Hoque2013}
A.~K. M.~M. {Hoque}, S.~O. {Amin}, A.~{Alyyan}, B.~{Zhang}, L.~{Zhang}, and
  L.~{Wang}, ``{NLSR: Named-data Link State Routing Protocol},'' in
  \emph{Proceedings of the 3rd ACM SIGCOMM Workshop on Information-centric
  Networking}, ser. ICN '13.\hskip 1em plus 0.5em minus 0.4em\relax New York,
  NY, USA: ACM, 2013, pp. 15--20.

\bibitem{Lehman2016a}
V.~{Lehman}, A.~{Gawande}, B.~{Zhang}, L.~{Zhang}, R.~{Aldecoa}, D.~{Krioukov},
  and L.~{Wang}, ``{An Experimental Investigation of Hyperbolic Routing with a
  Smart Forwarding Plane in NDN},'' in \emph{2016 IEEE/ACM 24th International
  Symposium on Quality of Service (IWQoS)}, June 2016, pp. 1--10.

\bibitem{Fu2017}
B.~{Fu}, L.~{Qian}, Y.~{Zhu}, and L.~{Wang}, ``{Reinforcement Learning-Based
  Algorithm for Efficient and Adaptive Forwarding in Named Data Networking},''
  in \emph{2017 IEEE/CIC International Conference on Communications in China
  (ICCC)}, Oct 2017, pp. 1--6.

\bibitem{Mnih2013}
V.~Mnih, K.~Kavukcuoglu, D.~Silver, A.~Graves, I.~Antonoglou, D.~Wierstra, and
  M.~Riedmiller, ``{Playing Atari with Deep Reinforcement Learning},''
  \emph{{arXiv preprint arXiv:1312.5602}}, 2013.

\bibitem{Afanasyev2014}
A.~Afanasyev, J.~Shi, B.~Zhang, L.~Zhang, I.~Moiseenko, Y.~Yu, W.~Shang,
  Y.~Huang, J.~P. Abraham, S.~DiBenedetto \emph{et~al.}, ``{NFD Developer's
  Guide},'' \emph{Dept. Comput. Sci., Univ. California, Los Angeles, Los
  Angeles, CA, USA, Tech. Rep. NDN-0021}, 2014.

\bibitem{Abadi2016}
M.~Abadi, A.~Agarwal, P.~Barham, E.~Brevdo, Z.~Chen, C.~Citro, G.~S. Corrado,
  A.~Davis, J.~Dean, M.~Devin, S.~Ghemawat, I.~Goodfellow, A.~Harp, G.~Irving,
  M.~Isard, Y.~Jia, R.~Jozefowicz, L.~Kaiser, M.~Kudlur, J.~Levenberg, D.~Mane,
  R.~Monga, S.~Moore, D.~Murray, C.~Olah, M.~Schuster, J.~Shlens, B.~Steiner,
  I.~Sutskever, K.~Talwar, P.~Tucker, V.~Vanhoucke, V.~Vasudevan, F.~Viegas,
  O.~Vinyals, P.~Warden, M.~Wattenberg, M.~Wicke, Y.~Yu, and X.~Zheng,
  ``{TensorFlow: Large-Scale Machine Learning on Heterogeneous Distributed
  Systems},'' \emph{{arXiv preprint arXiv:1603.04467}}, 2016.

\bibitem{Mastorakis2017}
S.~Mastorakis, A.~Afanasyev, and L.~Zhang, ``{On the Evolution of ndnSIM: an
  Open-Source Simulator for NDN Experimentation},'' \emph{SIGCOMM Comput.
  Commun. Rev.}, vol.~47, no.~3, pp. 19--33, Sep. 2017.

\bibitem{Yi2013}
C.~Yi, A.~Afanasyev, I.~Moiseenko, L.~Wang, B.~Zhang, and L.~Zhang, ``{A case
  for stateful forwarding plane},'' \emph{Computer Communications}, vol.~36,
  no.~7, pp. 779--791, 2013.

\end{thebibliography}

\end{document}